\documentclass[aps,showpacs,twocolumn,preprintnumbers,bibnotes]{revtex4-1}
\usepackage{ifpdf}
\usepackage{graphicx}
\usepackage{color}
\usepackage{dcolumn}
\usepackage{bm}
\usepackage{mathrsfs}
\usepackage{amsfonts}
\usepackage{amssymb}
\usepackage{amsmath}
\usepackage{dsfont}
\usepackage{longtable}
\usepackage{fixmath}
\usepackage{upgreek}
\usepackage{latexsym}
\usepackage[T1]{fontenc}% Has bold small caps - relevant to \textsc in sec. headings
\usepackage{ae,aecompl}
\usepackage{url}
\usepackage[dvipsnames*,svgnames]{xcolor}
\usepackage[pdfauthor={Behnam Farid},%
            pdftitle={Comment on ``Absence of Luttinger's Theorem'', by Kiaran B. Dave, Philip W. Phillips and Charles L. Kane, arXiv:1207.4201.},%
            plainpages=false,pdfpagelabels,pagebackref,%
            breaklinks=true,%
            hyperfootnotes=true,%
            bookmarks=true]{hyperref}
\hypersetup{colorlinks=true,%
            citecolor=blue,%DarkBlue,%
            filecolor=black,%
            linkcolor=DarkBlue,%
            urlcolor=blue,%
            pdftex,%
            pdfstartview={FitH}}
\usepackage[all]{hypcap}
\def\t#1{\tilde{#1}}
\def\wh#1{\widehat{#1}}
\def\h#1{\hat{#1}}
\def\b#1{\bar{#1}}

\def\tJ{$t$-$J$ }
\DeclareMathOperator{\re}{Re}
\DeclareMathOperator{\im}{Im}
\DeclareMathOperator{\e}{e}

\begin{document}
\ifx\href\undefined\else\hypersetup{linktocpage=true}\fi
% You should use BibTeX and apsrev.bst for references
\bibliographystyle{apsrev}
%\preprint{LT-2012-1}
\title{Comment on ``Absence of Luttinger's Theorem'', by Kiaran B. Dave, Philip W. Phillips and Charles L. Kane, \href{http://lanl.arxiv.org/abs/1207.4201}{\textsl{arXiv:1207.4201}}}

\author{Behnam Farid}
\email{behnam.farid@btinternet.com}
%\affiliation{}
%\date{August 12, 2012}
\date{\today}

\begin{abstract}
The explicit expression for the Luttinger number $N_{\textsc{l}}$ corresponding to the $\textrm{SU}(N)$ model of Dave, Phillips and Kane reveals that $N_{\textsc{l}}$ is equal to the number of particles $n$ only when $n = N/2$ (for $N$ even and assuming that the unit-step function $\Theta(x)$ is equal to $\frac{1}{2}$ at $x=0$) and $n = N$, signalling failure of the Luttinger theorem for all other values of $n$ in the interval $\{1,2,\dots,N\}$. In this Comment, we first present general arguments showing that the absence of the Luttinger theorem for the $\textrm{SU}(N)$ model under consideration is rooted in the non-uniqueness of the ground state of this model for $0 < n < N$, the validity of the Luttinger theorem for $n = N/2$, when $N$ even, being accidental, a consequence of particle-hole symmetry. Consequently, by supplementing the Hamiltonian of the $\textrm{SU}(N)$ model with a perturbation Hamiltonian that removes the ground-state degeneracy, the Luttinger theorem is to apply for the resulting model in the limit of the coupling constant $\lambda$ of this perturbation approaching zero, where the limit $\lambda \to 0$ is clearly to be taken \textsl{subsequent} to taking the zero-temperature limit of the thermal single-particle Green function in the expression for $N_{\textsc{l}}$. We explicitly establish the validity of this statement for the case of $N=4$. The details of the relevant calculations being distinctly transparent, one can readily convince oneself that our observation is valid for arbitrary $N$. It follows that the issues raised by Dave, Phillips and Kane, such as non-existence of the Luttinger-Ward functional and ``breakdown of the elemental particle picture in strongly correlated electron matter'', are all inessential to the observed failure of the Luttinger theorem. As regards the singularity of the self-energy $\Sigma(\omega)$ on the real $\omega$-axis, observed by Dave, Phillips and Kane, we demonstrate that this also is a direct consequence of the non-uniqueness of the ground state of the $\textrm{SU}(N)$ model for $0 < n < N$. In the light of the above observations, we are in a position to state that to this date \textsl{no} case has come to light  indicative of the failure of the Luttinger theorem under the conditions for which it has been deduced.
\end{abstract}

\maketitle
%{\scriptsize{\tableofcontents}}
{\footnotesize{\tableofcontents}}
% I:
\section{Introduction}
\label{s1}
In a recent publication, Dave, Phillips and Kane (hereafter DPK) \cite{DPK12} have considered the many-particle system described by the Hamiltonian
\begin{equation}
\wh{\mathcal{H}} = \frac{U}{2} (\h{n})^2,
\label{e1}
\end{equation}
where $U > 0$ is the interaction-energy parameter, and
\begin{equation}
\h{n} \equiv \sum_{\upalpha = 1}^N \h{n}_{\upalpha},\;\; \text{where}\;\; \h{n}_{\upalpha} \doteq \h{c}_{\upalpha}^{\dag} \h{c}_{\upalpha},
\label{e2}
\end{equation}
in which $\{ \h{c}_{\upalpha}\}$ are canonical annihilation fermion operators, and $\{\h{c}_{\upalpha}^{\dag}\}$ their Hermitian conjugates.

The thermal single-particle Green function $\mathscr{G}_{\upalpha\upalpha'}(\omega)$ pertaining to the $\textrm{SU}(N)$ model under discussion proves to be diagonal with respect to the indices $\upalpha, \upalpha'$ \cite{DPK12}. Consequently, with $G_{\upalpha\upalpha'}(\omega)$ denoting the zero-temperature limit (corresponding to $\beta \equiv 1/T \to \infty$ -- throughout this Comment $k_{\textsc{b}} = 1$) of the $\mathscr{G}_{\upalpha\upalpha'}(\omega)$ specific to the zero-temperature limit of the chemical potential $\mu_{\beta}$ corresponding to the mean value of particles $n$ in the grand-canonical ensemble (see Sec.~\ref{s2b1} however), the Luttinger number (LN) $N_{\textsc{l}}$ \cite[\S2]{BF07a} is defined as follows \cite[Eq.~(95)]{JML60}:
\begin{equation}
N_{\textsc{l}} \doteq \sum_{\upalpha=1}^{N} \Theta(G_{\upalpha\upalpha}(0)).
\label{e3}
\end{equation}
According to the Luttinger theorem (LT) \cite{LW60,JML60,BF07a}, one must have
\begin{equation}
N_{\textsc{l}} = n.
\label{e4}
\end{equation}
For the $\textrm{SU}(N)$ model, DPK \cite{DPK12} have however deduced that
\begin{equation}
N_{\textsc{l}} = N \Theta(2 n - N).
\label{e5}
\end{equation}
Assuming that $\Theta(0) = \frac{1}{2}$ (see however Ref.~\cite[\S2.4]{BF07a}), one observes that unless $n = N/2$ (for $N$ even) or $n=N$, the LT is clearly violated for the $\textrm{SU}(N)$ model under consideration \cite{DPK12}. In this Comment we clarify the reason underlying this failure. Briefly, by the non-uniqueness of the ground state (GS) of the $\textrm{SU}(N)$ model for $0< n < N$, this model is for $0 < n < N$ \emph{a priori} excluded from the set of models to which the LT is applicable. That despite the apparent non-uniqueness of the GS of this model for $0< n < N$ the LT proves to be valid for $N$ \textsl{even} and $n = N/2$, is a direct consequence of the explicit particle-hole symmetry of the problem at hand when in this case the chemical potential $\mu$ is identified with the zero-temperature limit of $\mu_{\beta}$.

% II:
\section{Analysis}
\label{s2}
% II.A:
\subsection{General considerations}
\label{s2a}
Let $\vert\Psi_{0,1}\rangle \equiv \vert 0, 0, \dots, 0\rangle$ ($N$ zeros) denote the normalized vacuum state of the $\textrm{SU}(N)$ model under consideration. For the normalized $n$-particle eigenstate $\vert \Psi_{n,s}\rangle$ of $\wh{\mathcal{H}}$, with $n\ge 1$, one has:
\begin{equation}
\vert \Psi_{n, s}\rangle = \Pi_{j=1}^n \h{c}_{\nu_s(j)}^{\dag} \vert\Psi_{0,1}\rangle,\;\; s \in\big\{ 1,2, \dots, \binom{N}{n}\big\},
\label{e6}
\end{equation}
where the integer-valued mapping
\begin{equation}
\nu_s: \{1,\dots, n\} \mapsto \{1,\dots, N\},
\label{e7}
\end{equation}
for a given $s$, coincides with an $N$-permutation $\mathcal{P}_{N,s'}$ of the ordered set $\{1, 2, \dots, N\}$. Here, the two permutations $\mathcal{P}_{N,s'}$ and $\mathcal{P}_{N,s''}$ are to be identified when the ordered set $\{ \mathcal{P}_{N,s'} 1, \dots, \mathcal{P}_{N,s'} n\}$ is an $n$-permutation of the ordered set $\{ \mathcal{P}_{N,s''} 1, \dots, \mathcal{P}_{N,s''} n\}$. Equivalently as regards the ordered sets $\{ \mathcal{P}_{N,s'} (n+1), \dots, \mathcal{P}_{N,s'} N\}$ and $\{ \mathcal{P}_{N,s''} (n+1), \dots, \mathcal{P}_{N,s''} N\}$. Hence, one is left with $N!/(n! (N-n)!) \equiv \binom{N}{n}$ distinct permutations, a fact reflected in the specific set over which the integer $s$ in Eq.~(\ref{e6}) varies. Note that because of the anti-commutation relation $[\h{c}_{\upalpha\phantom{'}}^{\dag}, \h{c}_{\upalpha'}^{\dag}]_- = 0$, $\forall\upalpha,\upalpha'$, identifying $\nu_s$ with either $\mathcal{P}_{N,s'}$ or $\mathcal{P}_{N,s''}$, with $\{ \mathcal{P}_{N,s'} 1, \dots, \mathcal{P}_{N,s'} n\}$ and $\{ \mathcal{P}_{N,s''} 1, \dots, \mathcal{P}_{N,s''} n\}$ differing by an $n$-permutation, the corresponding $\vert \Psi_{n, s}\rangle$ differ by at most a minus sign.

For later reference, the state $\vert\Psi_{n, s}\rangle$, Eq.~(\ref{e6}), can be represented as follows (compare with the $\vert\Psi_{0,1}\rangle$ introduced above):
\begin{equation}
\vert \Psi_{n, s}\rangle = \vert n_{s,1}, n_{s,2},\dots, n_{s,N}\rangle,
\label{e8}
\end{equation}
where $n_{s,\upalpha} \in \{0,1\}$, $\forall\upalpha$, and
\begin{equation}
\sum_{\upalpha =1}^N n_{s,\upalpha} = n,\; \forall s \in\big\{1,2,\dots, \binom{N}{n}\big\}.
\label{e9}
\end{equation}

From
\begin{equation}
\wh{\mathcal{H}}\hspace{0.6pt} \vert \Psi_{n,s}\rangle = \frac{U}{2} n^2\hspace{0.6pt} \vert \Psi_{n, s}\rangle, \;\forall s,
\label{e10}
\end{equation}
it is observed that for $n$ particles the energy level $U n^2/2$ of the $\textrm{SU}(N)$ model is $\binom{N}{n}$-fold degenerate. For later reference, we point out that the GS energy of the Hubbard Hamiltonian \cite{PWA59,JH63} for $n$ spin-$\frac{1}{2}$ particles defined on a lattice comprised of $N$ sites is in the atomic limit \cite[\S5]{JH63} $2^n \binom{N}{n}$-fold degenerate for $n \le N$ \cite[\S4.4.2]{PF03}, where the factor $2^n$ corresponds to the spin degeneracy; at half-filling, corresponding to $n = N$, one is thus left only with spin degeneracy. For definiteness, with $\{T_{i,j}\}$ denoting the hopping integrals in the Hubbard Hamiltonian, the atomic limit is defined by the condition $T_{i,j} = 0$ for \textsl{all} $i\not= j$ \cite{JH63}. The term $T_{i,i} \equiv T_0$ is arbitrary and is usually absorbed in the chemical potential $\mu$ (cf. Eq.~(\ref{e17})), however it has been explicitly taken into account in Ref.~\cite{JH63}.

Two main problems arise as a result of the degeneracy of the GS energy of the $\textrm{SU}(N)$ model for $n \not= 0, N$.

First, in the cases where the GS is not unique, for $\mu \in (\mu_n^-,\mu_n^+)$, Eq.~(\ref{e33}) \cite[Eq.~(B.7)]{BF07a}, the zero-temperature limit of the thermal single-particle Green function coincides with an \textsl{average} over the set of zero-temperature single-particle Green functions each member of which corresponds to one of the GSs in $\{\vert\Psi_{n,s}\rangle \| s\}$. This average function has \emph{no} place in the formulation of the LT. In Ref.~\cite[Appendix C]{BF07a} we have at places indicated the consequence of degenerate GS energies (see for instance Eq.~(C.22) and the remark following Eq.~(C.23) in Ref.~\cite{BF07a}; the expression in Eq.~(C.22) is to be viewed in the light of the expressions in Eqs.~(C.12), (C.19) -- (C.21)). Later, in Sec.~\ref{s2b2}, we explicitly show how the zero-temperature single-particle Green function obtained through the thermal averaging over the manifold of the single-particle Green functions corresponding to the $n$-particle GSs $\{ \vert\Psi_{n,s}\rangle \| s\}$ in general yields an incorrect (from the perspective of the LT) LN $N_{\textsc{l}}$, Eq.~(\ref{e3}).

Second, the LT has been explicitly proved within the framework of the time-dependent many-body perturbation theory \cite{FW03} (for a comprehensive discussion, see in particular Ref.~\cite{BF07a}). In this framework, it is \textsl{assumed} that at zero temperature an adiabatic increase of the coupling constant of the interaction Hamiltonian, from zero at $\uptau=-\infty$ to the full strength at $\uptau=0$, results in an adiabatic evolution of the unique (up to a trivial phase factor) $n$-particle non-interacting GS into the unique $n$-particle interacting GS. The latter state is further \textsl{assumed} adiabatically to evolve, on adiabatically decreasing the strength of interaction to zero, into an $n$-particle state at $\uptau=\infty$ that up to a possible trivial phase factor coincides with the non-interacting $n$-particle GS with which one has begun at $\uptau=-\infty$. According to the celebrated Gell-Mann and Low theorem \cite{GL51}, on which the time-dependent many-body perturbation theory is based, the first of the above-mentioned adiabatic processes in general generates an $n$-particle \textsl{eigenstate} of the full Hamiltonian at $\uptau=0$, not necessarily its $n$-particle GS \cite[p.~61]{FW03}. Failure of the state at $\uptau=0$ to be the interacting GS, amounts to a failure of the many-body perturbation expansion \cite{BF99} (see also in particular the remarks in \S5.1.1 of Ref.~\cite{BF07a}).

For completeness, as demonstrated in Ref.~\cite{BF07a}, for a class of systems, fully described in the latter reference, many-body perturbation theory -- in the specific way it has been employed in the proof of the LT \cite{LW60}, does \textsl{not} break down. Briefly, expansion in terms of the \textsl{skeleton} self-energy diagrams \cite{LW60} \cite[\S5.3.1]{BF07a} evaluated in terms of the \textsl{exact} interacting single-particle Green function safeguards the relevant perturbation series against either diverging or converging to false limits.

The LT having originally been implicitly deduced for metallic GSs, on account of an earlier observation by Rosch \cite{AR07a}, in Ref.~\cite[\S6.1]{BF07a} (see also Ref.~\cite{BF07b}) we have demonstrated that a possible failure of the LT for non-metallic GSs, when the chemical potential $\mu \in (\mu_n^-,\mu_n^+)$ differs from the zero-temperature limit of $\mu_{\beta}$, is \textsl{not} due to a possible failure of the many-body perturbation theory, but due to the possibility of arriving at a \textsl{false limit} \cite[\S302-306]{EWH50} for $N_{\textsc{l}}$ in the process of effecting the zero-temperature limit $\beta\to\infty$ for any fixed value of $\mu$ inside $(\mu_n^-,\mu_n^+)$ different from $\mu_{\infty}$; we have also demonstrated that the last-mentioned \textsl{false limit} is similarly avoided by identifying $\mu$ with $\mu_{\beta}$ in the process of effecting the zero-temperature limit \cite{Note1}. We note in passing that in Ref.~\cite{FT09} we have rigorously dealt with the half-filled GS of the one-dimensional $t$-$t'$-$V$ model, which for $V$ greater than a critical value $V_{\textrm{c}}(t,t')$ is a charge-density-wave state and insulating. For this insulating GS, the LT proves to be trivially satisfied for \textsl{all} $\mu$ inside the relevant gap $(\mu_n^-,\mu_n^+)$ \cite[\S\hspace{0.5pt}IV]{FT09} (see Sec.~\ref{sac1} in appendix \ref{sac}).

% II.B:
\subsection{Explicit treatment of the \texorpdfstring{$\textrm{SU}(N)$}{} model for \texorpdfstring{$N=4$}{}}
\label{s2b}
% II.B.1:
\subsubsection{Removal of the GS degeneracy}
\label{s2b1}
Below we explicitly consider the specific case of $N=4$, for which the relevant Fock space is spanned by the following set of $\sum_{n=0}^4 \binom{4}{n} = 2^4$ simultaneous eigenstates of $\wh{\mathcal{H}}$ and $\h{n}$ (cf. Eqs.~(\ref{e8}) and (\ref{e9})):
\begin{equation}
\vert \Psi_{0, 1}\rangle = \vert 0, 0, 0, 0 \rangle,
\label{e11}
\end{equation}
\begin{eqnarray}
\vert \Psi_{1, 1}\rangle &=& \vert 1, 0, 0, 0\rangle,\; \vert \Psi_{1, 2}\rangle = \vert 0, 1, 0, 0\rangle, \nonumber\\
\vert \Psi_{1, 3}\rangle &=& \vert 0, 0, 1, 0\rangle,\; \vert \Psi_{1, 4}\rangle = \vert 0, 0, 0, 1\rangle,
\label{e12}
\end{eqnarray}
\begin{eqnarray}
\vert \Psi_{2, 1}\rangle &=& \vert 1, 1, 0, 0\rangle,\; \vert \Psi_{2, 2}\rangle = \vert 1, 0, 1, 0\rangle, \nonumber\\
\vert \Psi_{2, 3}\rangle &=& \vert 1, 0, 0, 1\rangle,\; \vert \Psi_{2, 4}\rangle = \vert 0, 1, 1, 0\rangle, \nonumber\\
\vert \Psi_{2, 5}\rangle &=& \vert 0, 1, 0, 1\rangle,\; \vert \Psi_{2, 6}\rangle = \vert 0, 0, 1, 1\rangle,
\label{e13}
\end{eqnarray}
\begin{eqnarray}
\vert \Psi_{3, 1}\rangle &=& \vert 1, 1, 1, 0\rangle,\; \vert \Psi_{3, 2}\rangle = \vert 1, 1, 0, 1\rangle, \nonumber\\
\vert \Psi_{3, 3}\rangle &=& \vert 1, 0, 1, 1\rangle,\; \vert \Psi_{3, 4}\rangle = \vert 0, 1, 1, 1\rangle,
\label{e14}
\end{eqnarray}
\begin{equation}
\vert \Psi_{4, 1}\rangle = \vert 1, 1, 1, 1\rangle.
\label{e15}
\end{equation}
In assigning an $s$ to a particular state corresponding to a given $n$, where $s\in \{1,2,\dots,\binom{4}{n}\}$, we have had in mind the following `perturbed' Hamiltonian (cf. Eq.~(\ref{e2})):
\begin{equation}
\wh{\mathcal{H}}_{\lambda} \doteq \wh{\mathcal{H}} + \lambda \sum_{\upalpha=1}^{N} \upalpha\hspace{0.6pt} \h{n}_{\upalpha}.
\label{e16}
\end{equation}
Two significant properties of this Hamiltonian are: firstly, that for all $\lambda \in \mathds{R}$ its eigenstates coincide with those of $\wh{\mathcal{H}}$ and $\h{n}$, presented in Eq.~(\ref{e6}), and, secondly, that its GS is unique for \emph{all} $\lambda > 0$. With reference to the Hamiltonian in Eq.~(\ref{e16}), the energies of the states in Eqs.~(\ref{e11}) -- (\ref{e15}) are non-decreasing for increasing values of both $n$ and $s$. Thus, for any $n \in \{1,2,3,4\}$, $\vert\Psi_{n,1}\rangle$ is the $n$-particle GS of $\wh{\mathcal{H}}_{\lambda}$ for \textsl{all} $\lambda >0$.

Let $\{K_{s}(n,\lambda)\}$ denote the eigenvalues corresponding to eigenstates $\{\vert\Psi_{n, s}\rangle\}$ of the thermodynamic Hamiltonian
\begin{equation}
\wh{\mathcal{K}}_{\lambda} \doteq \wh{\mathcal{H}}_{\lambda} - \mu \h{n},
\label{e17}
\end{equation}
where $\mu$ is the chemical potential. For the corresponding thermal single-particle Green function $\mathscr{G}_{\upalpha\upalpha'}(\omega,\lambda)$, one has (throughout this Comment, $\hbar=1$) \cite[Eq.~(C.19)]{BF07a}:
\begin{equation}
\mathscr{G}_{\upalpha\upalpha'}(\omega,\lambda) = \frac{1}{\mathcal{Z}(\lambda)} \sum_{n=0}^{N} \sum_{s \in \mathfrak{S}(n)}\hspace{-0.5pt} \e^{-\beta K_s(n,\lambda)}\hspace{-0.5pt} f_{\upalpha\upalpha'}(\omega,\lambda;n,s),
\label{e18}
\end{equation}
where \cite[Eq.~(C.9)]{BF07a}
\begin{equation}
\mathcal{Z}(\lambda) = \sum_{n=0}^{N} \sum_{s\in \mathfrak{S}(n)} \e^{-\beta K_s(n,\lambda)}
\label{e19}
\end{equation}
is the grand partition function,
\begin{equation}
\mathfrak{S}(n) \doteq \left\{\begin{array}{ll} \{1,\dots, \binom{N}{n}\}, & n \le N, \\ \\
\varnothing, & n > N,\\ \end{array} \right.
\label{e20}
\end{equation}
where $\varnothing$ denotes the empty set, and \cite[Eq.~(C.20)]{BF07a}
\begin{widetext}
\begin{equation}
f_{\upalpha\upalpha'}(\omega,\lambda;n,s) \doteq \sum_{s'\in \mathfrak{S}(n-1)} \frac{(\mathbb{A}_{\upalpha'}(n))_{s,s'} (\mathbb{A}_{\upalpha}^{\dag}(n))_{s',s}}{\omega - K_s(n,\lambda) + K_{s'}(n-1,\lambda)} + \sum_{s' \in \mathfrak{S}(n+1)} \frac{ (\mathbb{A}_{\upalpha}^{\dag}(n+1))_{s,s'} (\mathbb{A}_{\upalpha'}(n+1))_{s',s} }{\omega - K_{s'}(n+1,\lambda) + K_{s}(n,\lambda)},
\label{e21}
\end{equation}
\end{widetext}
in which
\begin{equation}
(\mathbb{A}_{\upalpha}(n))_{s,s'} \doteq \langle \Psi_{n, s}\vert \h{c}_{\upalpha}^{\dag} \vert \Psi_{n -1, s'}\rangle.
\label{e22}
\end{equation}
For the states $\{\vert \Psi_{n, s}\rangle\}$ given in Eqs.~(\ref{e11}) - (\ref{e15}), one readily obtains (below, $\b{\delta}_{\upalpha,\upalpha'} \equiv -\delta_{\upalpha,\upalpha'}$):
\begin{equation}
\mathbb{A}_{\upalpha}(1) =
\begin{pmatrix}
\delta_{\upalpha,1} \\
\delta_{\upalpha,2} \\
\delta_{\upalpha,3} \\
\delta_{\upalpha,4}
\end{pmatrix}\hspace{-2.0pt},\;
\mathbb{A}_{\upalpha}(2) =
\begin{pmatrix}
\b{\delta}_{\upalpha,2} & \delta_{\upalpha,1} & 0 & 0 \\
\b{\delta}_{\upalpha,3} & 0 & \delta_{\upalpha,1} & 0 \\
\b{\delta}_{\upalpha,4} & 0 & 0 & \delta_{\upalpha,1} \\
0 & \b{\delta}_{\upalpha,3} & \delta_{\upalpha,2} & 0 \\
0 & \b{\delta}_{\upalpha,4} & 0 & \delta_{\upalpha,2} \\
0 & 0 & \b{\delta}_{\upalpha,4} & \delta_{\upalpha,3}
\end{pmatrix}\hspace{-2.0pt},
\label{e23}
\end{equation}
\begin{equation}
\mathbb{A}_{\upalpha}(3) =
\begin{pmatrix}
\delta_{\upalpha,3} & \b{\delta}_{\upalpha,2} & 0 & \delta_{\upalpha,1} & 0 & 0 \\
\delta_{\upalpha,4} & 0 & \b{\delta}_{\upalpha,2} & 0 & \delta_{\upalpha,1} & 0 \\
0 & \delta_{\upalpha,4} & \b{\delta}_{\upalpha,3} & 0 & 0 & \delta_{\upalpha,1} \\
0 & 0 & 0 & \delta_{\upalpha,4} & \b{\delta}_{\upalpha,3} & \delta_{\upalpha,2}
\end{pmatrix}\hspace{-2.0pt},
\label{e24}
\end{equation}
\begin{equation}
\mathbb{A}_{\upalpha}(4) =
\begin{pmatrix}
\b{\delta}_{\upalpha,4} & \delta_{\upalpha,3} & \b{\delta}_{\upalpha,2} & \delta_{\upalpha,1}
\end{pmatrix}\hspace{-2.0pt}.
\label{e25}
\end{equation}
For the considerations of the present Comment, we need to calculate the function
\begin{equation}
G_{\upalpha\upalpha'}(\omega,\lambda) \equiv \lim_{\beta\to\infty} \mathscr{G}_{\upalpha\upalpha'}(\omega,\lambda),\; \lambda > 0,
\label{e26}
\end{equation}
corresponding to the $n$-particle GS of $\wh{\mathcal{H}}_{\lambda}$, $\lambda >0$. From the above expressions, one readily deduces that $\mathscr{G}_{\upalpha\upalpha'}(\omega,\lambda)$ is identically vanishing for $\upalpha \not=\upalpha'$, $\forall\beta, \lambda$, similar to the case of $\lambda=0$ considered in Ref.~\cite{DPK12}.

Since the $n$-particle GS of $\wh{\mathcal{H}}_{\lambda}$ is unique for all $\lambda >0$, equal to $\vert\Psi_{n,1}\rangle$, following the asymptotic expression in Eq.~(C.12) in Ref.~\cite{BF07a} for $\mu \in (\mu_n^-,\mu_n^+)$, one immediately obtains
\begin{eqnarray}
&&\hspace{-0.7cm} G_{\upalpha\upalpha}(\omega,\lambda) = \sum_{s \in \mathfrak{S}(n-1)}  \frac{\vert (\mathbb{A}_{\upalpha}(n))_{1,s}\vert^2}{\omega - K_1(n,\lambda) + K_s(n-1,\lambda)} \nonumber\\
&&\hspace{1.0cm} + \sum_{s \in \mathfrak{S}(n+1)}  \frac{\vert (\mathbb{A}_{\upalpha}(n+1))_{s,1}\vert^2}{\omega - K_s(n+1,\lambda) + K_1(n,\lambda)}.
\label{e27}
\end{eqnarray}
Making use of the representation in Eq.~(\ref{e27}) and the above expressions for the matrices $\{\mathbb{A}_{\upalpha}(n) \| n\}$, one readily arrives at the following results:
\begin{eqnarray}
&&\hspace{-1.0cm} \left. G_{\upalpha\upalpha}(0,\lambda)\right|_{n=1} = \frac{\delta_{\upalpha,1}}{\mu - U/2 -\lambda} + \frac{\delta_{\upalpha,2}}{\mu - 3 U/2 - 2\lambda}\nonumber\\
&&\hspace{1.0cm}+ \frac{\delta_{\upalpha,3}}{\mu - 3 U/2 -3\lambda} + \frac{\delta_{\upalpha,4}}{\mu - 3 U/2 - 4\lambda},
\label{e28}
\end{eqnarray}
\begin{eqnarray}
&&\hspace{-1.0cm} \left. G_{\upalpha\upalpha}(0,\lambda)\right|_{n=2} = \frac{\delta_{\upalpha,1}}{\mu - 3 U/2 -\lambda} + \frac{\delta_{\upalpha,2}}{\mu-3 U/2 -2\lambda}\nonumber\\
&&\hspace{1.2cm}+ \frac{\delta_{\upalpha,3}}{\mu -5 U/2 -3\lambda} + \frac{\delta_{\upalpha,4}}{\mu - 5 U/2 - 4\lambda},
\label{e29}
\end{eqnarray}
\begin{eqnarray}
&&\hspace{-1.0cm} \left. G_{\upalpha\upalpha}(0,\lambda)\right|_{n=3} = \frac{\delta_{\upalpha,1}}{\mu - 5 U/2 -\lambda} + \frac{\delta_{\upalpha,2}}{\mu-5 U/2 -2\lambda}\nonumber\\
&&\hspace{1.2cm}+ \frac{\delta_{\upalpha,3}}{\mu -5 U/2 -3\lambda} + \frac{\delta_{\upalpha,4}}{\mu - 7 U/2 - 4\lambda},
\label{e30}
\end{eqnarray}
\begin{eqnarray}
&&\hspace{-0.8cm} \left. G_{\upalpha\upalpha}(0,\lambda)\right|_{n=4} = \frac{\delta_{\upalpha,1}}{\mu - 7 U/2 -\lambda} + \frac{\delta_{\upalpha,2}}{\mu-7 U/2 -2\lambda}\nonumber\\
&&\hspace{1.4cm}+ \frac{\delta_{\upalpha,3}}{\mu -7 U/2 -3\lambda} + \frac{\delta_{\upalpha,4}}{\mu - 7 U/2 - 4\lambda}.
\label{e31}
\end{eqnarray}

We introduce \cite[Eqs.~(B.5) and (B.6)]{BF07a}
\begin{eqnarray}
\mu_n^- &\doteq& H(n) - H(n-1) = (n-\frac{1}{2})\hspace{0.1pt} U,\;\, 0 < n \le N, \nonumber\\
\mu_n^+ &\doteq& H(n+1) - H(n) = \left\{\begin{array}{ll}\displaystyle  (n+\frac{1}{2})\hspace{0.1pt} U, & n < N,\\ \\ \infty, & n = N, \end{array} \right.
\label{e32}
\end{eqnarray}
where $H(n) \doteq U n^2/2$ is the degenerate eigenvalue of $\wh{\mathcal{H}}$ corresponding to $\vert\Psi_{n,s}\rangle$, $s \in \{1,\dots, \binom{N}{n}\}$, Eq.~(\ref{e10}).  Note in passing that in contrast to Ref.~\cite{DPK12}, here we do not consider $H(n+1)$ as being a definite quantity for $n=N$ (cf. Eq.~(\ref{e20})). It is observed that in the limit of $\lambda =0^+$ for \cite[Eq.~(B.7)]{BF07a}
\begin{equation}
\mu \in (\mu_n^-,\mu_n^+)
\label{e33}
\end{equation}
the denominators of the first $n$ terms on the right-hand sides of the above expressions for $G_{\upalpha\upalpha}(0,\lambda)$, with $\lambda=0^+$, are positive. For $N=4$, one thus has the following \emph{exact} result:
\begin{equation}
\sum_{\upalpha=1}^N \Theta\big(G_{\upalpha\upalpha}(0,0^+)\big) = n,\;\; \forall \mu \in (\mu_n^-,\mu_n^+),
\label{e34}
\end{equation}
implying the validity of the LT for \textsl{all} $n \in \{1, 2, 3, 4\}$, not only for $\mu$ identified with the zero-temperature limit $\mu_{\infty}$ of the chemical potential $\mu_{\beta}$ corresponding to $n$ particles  (which in the case at hand is equal to $U n$ \cite{DPK12} for $n < N$, and not less than $U n$ for $n=N$), but also for \textsl{all} $\mu$ in the single-particle excitation gap $(\mu_n^-,\mu_n^+)$ of the $n$-particle GS of the model under consideration, with $\lambda = 0^+$. The equality in Eq.~(\ref{e34}) is to be contrasted with that in Eq.~(\ref{e5}). \emph{We have therefore rigorously demonstrated that in the specific case of $N=4$ the failure of the LT as observed by DPK \cite{DPK12} is wholly attributable to the $n$-particle GS of $\wh{\mathcal{H}}_{\lambda=0}$, to be distinguished from $\wh{\mathcal{H}}_{\lambda=0^+}$, not being unique for $0< n < N$.}

% II.B.2
\subsubsection{A consequence of the thermal averaging of the Green functions corresponding to different GSs}
\label{s2b2}
In connection with the above observation, we note that by choosing the perturbation Hamiltonian in such a way that the state $\vert\Psi_{n,s_0}\rangle$, with $s_0 \in \{1,2,\dots,\binom{N}{n}\}\backslash \{1\}$, would be the GS of $\wh{\mathcal{H}}_{\lambda}$, we would have obtained expressions for $\lim_{\lambda\downarrow 0} \lim_{\beta\to\infty} \mathscr{G}_{\upalpha\upalpha}(0,\lambda)$ (note the order of the limits), corresponding to different values of $n$, similar to those given in Eqs.~(\ref{e28}) -- (\ref{e31}), \textsl{except} for a non-trivial permutation of the $\delta_{\upalpha,j}$, $j = 1,2,3,4$, in the numerators, which can be readily inferred from the expressions for the matrices $\{\mathbb{A}_{\upalpha}(n)\| n\}$ given in Eqs.~(\ref{e23}) -- (\ref{e25}) above. This is relevant, in that it makes explicit that the zero-temperature limit of the thermal single-particle Green function $\mathscr{G}_{\upalpha\upalpha'}(\omega)$,  corresponding to $\wh{\mathcal{H}}$ and $\mu \in (\mu_n^-,\mu_n^+)$, amounts to the \textsl{arithmetic mean} of $\{G_{\upalpha\upalpha'}^{(s)}(\omega)\| s\}$, $\binom{N}{n}$ single-particle zero-temperature Green functions, each corresponding to an $n$-particle GS of $\wh{\mathcal{H}}$ in $\{\vert\Psi_{n,s}\rangle \| s\}$ singled out through introducing an appropriate perturbation Hamiltonian and taking the limit of the coupling-constant $\lambda$ of this perturbation approaching zero subsequent to effecting the zero-temperature limit (see the remark following Eq.~(C.23) in Ref.~\cite{BF07a}). Thus, from the expressions in Eqs.~(\ref{e28}) -- (\ref{e31}) one trivially obtains that
\begin{eqnarray}
\lim_{\beta\to\infty} \left. \mathscr{G}_{\upalpha\upalpha}(0,0)\right|_{n=1} &=& \frac{1}{4} \Big\{ \frac{1}{\mu-U/2} + \frac{3}{\mu-3U/2}\Big\}, \hspace{0.6cm}
\label{e35}\\
\lim_{\beta\to\infty} \left. \mathscr{G}_{\upalpha\upalpha}(0,0)\right|_{n=2} &=& \frac{1}{6} \Big\{ \frac{3}{\mu-3U/2} + \frac{3}{\mu-5U/2}\Big\},
\label{e36}\\
\lim_{\beta\to\infty} \left. \mathscr{G}_{\upalpha\upalpha}(0,0)\right|_{n=3} &=& \frac{1}{4} \Big\{ \frac{3}{\mu-5U/2} + \frac{1}{\mu-7U/2}\Big\},
\label{e37}\\
\lim_{\beta\to\infty} \left. \mathscr{G}_{\upalpha\upalpha}(0,0)\right|_{n=4} &=& \frac{1}{\mu-7U/2},
\label{e38}
\end{eqnarray}
where the integers $1$ and $3$ in the numerators are in fact multipliers of $(\delta_{\upalpha,1} + \delta_{\upalpha,2} + \delta_{\upalpha,3} + \delta_{\upalpha,4}) \equiv 1$, the latter identity applying on account of $\upalpha \in \{1,2,3,4\}$ in the case of $N=4$.

By identifying the $\mu$ in the expressions in Eqs.~(\ref{e35}) -- (\ref{e38}) with $U n$, one readily obtains that
\begin{equation}
\lim_{\beta \to \infty} \lim_{\lambda\downarrow 0} \left.\mathscr{G}_{\upalpha\upalpha}(0,\lambda)\right|_{\mu= U n} = \frac{n-2}{U},\; n \in \{1,2,3,4\},
\label{e39}
\end{equation}
in full conformity with the expression in Eq.~(13) of Ref.~\cite{DPK12}, which is specific to $\mu = U n$, specialized to the case of $N=4$. For the case of $N=4$, we have thus explicitly demonstrated that
\begin{equation}
\lim_{\beta\to\infty} \lim_{\lambda\downarrow 0} \mathscr{G}_{\upalpha\upalpha}(\omega,\lambda) \not\equiv \lim_{\lambda\downarrow 0} \lim_{\beta\to\infty} \mathscr{G}_{\upalpha\upalpha}(\omega,\lambda),\;\; 0 < n < N.
\label{e40}
\end{equation}
One can easily convince oneself that this result is not specific to $N=4$, but applies for \textsl{all} $N$. The fact that despite this result the LT proves to apply for $\lim_{\beta\to\infty} \lim_{\lambda\downarrow 0} \mathscr{G}_{\upalpha\upalpha}(\omega,\lambda)$ when $N$ is even, $n=N/2$ and $\mu = (\mu_n^- + \mu_n^+)/2 \equiv U n$, is a direct consequence of particle-hole symmetry, whereby the latter zero-temperature single-particle Green function is equal to zero at $\omega = 0$ (cf. Eq.~(\ref{e39}), as well as Eqs.~(9) and (C242) in Ref.~\cite{FT09}). It is important to realize that each of the zero-temperature Green functions contributing to the function $\lim_{\beta\to\infty} \mathscr{G}_{\upalpha\upalpha}(0,0)$ in Eqs.~(\ref{e35}) -- (\ref{e38}) satisfies the requirement of the LT. In other words, \emph{with $N_{\textsc{l}}^{(s)}$ denoting the LN corresponding to $G_{\upalpha\upalpha'}^{(s)}(\omega)$, the zero-temperature single-particle Green function corresponding to the $n$-particle GS $\vert\Psi_{n,s}\rangle$, the thermal average of $\{ N_{\textsc{l}}^{(s)}\| s\}$, in the zero-temperature limit, satisfies the LT.}

% III:
\section{Summary and discussions}
\label{s3}
Above we have first indicated that in the cases where the $n$-particle GS of a system is not unique, the zero-temperature \textsl{limit} of the thermal single-particle Green function $\mathscr{G}_{\upalpha\upalpha'}(\omega)$ corresponding to a $\mu \in (\mu_n^-,\mu_n^+)$  has \textsl{no} place in the formulation of the LT \cite{BF07a}, it being an average (explicitly, arithmetic mean) of the zero-temperature Green functions $\{ G_{\upalpha\upalpha'}^{(s)}(\omega) \| s\}$ corresponding to the set of $n$-particle GSs $\{\vert\Psi_{n,s}\rangle \| s\}$. The failure of the LT for the $\textrm{SU}(N)$ model under discussion for a general $n$, first established by DPK \cite{DPK12}, is a direct manifestation of this fact. For the specific case of $N=4$, we have explicitly shown how removal of the degeneracy of the GS energy by an infinitesimal amount leads to full restoration the LT. The transparency of the underlying calculations lead one to conclude that this result is general, applying to arbitrary values of $N$. The validity of the LT in the case of $n = N/2$, for $N$ even, and the chemical potential $\mu$ equated with the zero-temperature limit of $\mu_{\beta}$ (assuming that $\Theta(0) = 1/2$), despite the degeneracy of the relevant GS energy, is accidental, a consequence of particle-hole symmetry.

From the experimental perspective, the GS energy of a system need not be degenerate in order for the LT to be observed as violated, as a consequence of the thermal energy $T$ not being sufficiently small with respect to the lowest-lying single-particle excitation energies; for insufficiently small $T$, the thermal single-particle Green function $\mathscr{G}_{\upalpha\upalpha'}(\omega)$ will necessarily contain considerable contributions corresponding to excited states, which, depending on the system under investigation, can prove destructive to the LT. For $N=4$ and a fixed value of $\lambda > 0$, we are able numerically to follow the development of $\mathscr{G}_{\upalpha\upalpha}(\omega,\lambda)$, $\forall\upalpha \in \{1,2,3,4\}$, as a function of $\beta$ and show how the LN $N_{\textsc{l}}$ as expressed in terms of $\mathscr{G}_{\upalpha\upalpha}(0,\lambda)$, corresponding to either a fixed value of $\mu \in (\mu_n^-,\mu_n^+)$ or $\mu=\mu_{\beta}$, deviates from $n$ for $\beta$ less than a critical value $\beta_{\textrm{c}}(\lambda)$, and how $N_{\textsc{l}}$ coincides exactly with $n$ for all $\beta > \beta_{\textrm{c}}(\lambda)$. \emph{We provide the Mathematica$^{\copyright}$ code underlying these calculations through an ancillary notebook file associated with the present text.}

We conclude this summary by stating that to our best knowledge to this date \textsl{no} case has come to light  indicative of the failure of the Luttinger theorem under the conditions for which it has been deduced.

% Appendices.
\begin{appendix}
% Appendices:
% A:
\section{Comments on some observations by Dave, Phillips and Kane \protect\cite{DPK12}}
\label{saa}
Some comments of general interest on some of the observations by DPK in Ref.~\cite{DPK12} are in place. For convenience of reference, we enumerate these comments.

\emph{(i)} The expression in Eq.~(22) of Ref.~\cite{DPK12}, with $I_1$ and $I_2$ as defined herein, is an \textsl{identity}, as emphasized in \S4 of Ref.~\cite{BF07a} (see in particular Eq.~(4.13) as well as \S6.2.3 of this reference). Consequently, reproducing the result $I_1 + I_2 = n$ solely signifies correctness of the underlying calculations, and \textsl{nothing} more.

\emph{(ii)} The singularity of the self-energy $\Sigma_{\upalpha\upalpha'}(\omega)$ along the \textsl{real} $\omega$ axis as deduced by DPK \cite{DPK12} (from the zero-temperature \textsl{limit} of the thermal single-particle Green function $\mathscr{G}_{\upalpha\upalpha'}(\omega)$ for $\mu \in (\mu_n^-,\mu_n^+)$), is a direct consequence of the non-uniqueness of the $n$-particle GS of the $\textrm{SU}(N)$ model under consideration for $0 < n < N$. Before elaborating on this statement, it is interesting to note that the self-energy $\Sigma(\omega)$ in Eq.~(17) of Ref.~\cite{DPK12} is identically vanishing for $n=0, N$, the values of $n$ for which the GS energy is non-degenerate. Here $\Sigma(\omega)$ is the short-hand notation for the diagonal element $\Sigma_{\upalpha\upalpha}(\omega)$ of the self-energy, which proves to be independent of $\upalpha$, in conformity with the expressions on the right-hand sides of Eqs.~(\ref{e35}) -- (\ref{e38}) in the main text of the present Comment.

Comparing the expressions in Eqs.~(\ref{e28}) -- (\ref{e31}) with their counterparts in Eqs.~(\ref{e35}) -- (\ref{e38}), one observes that the Green functions in the latter expressions, in contrast to those in the former ones, are independent of the index $\upalpha$, a fact fully attributable to the non-uniqueness of the $n$-particle GSs of the $\textrm{SU}(N)$ for $0 < n < N$. One readily verifies that the zero-temperature self-energy $\Sigma_{\upalpha\upalpha}(\omega,\lambda)$, $\forall\upalpha \in\{1,2,\dots,N\}$, associated with $G_{\upalpha\upalpha}(\omega,\lambda)$, $\lambda > 0$, and $0 < n < N$ depends not only non-trivially on $\upalpha$, but also is \textsl{independent} of $\omega$ \cite{Note2}. It is therefore analytic everywhere on the complex $\omega$ plane. The independence from $\upalpha$ of the $\Sigma_{\upalpha\upalpha}(\omega)$ as determined by DPK \cite{DPK12} and the simple pole of this function on the real $\omega$ axis are interdependent and follow directly from the non-uniqueness of the $n$-particle GS of the $\textrm{SU}(N)$ for $0 < n < N$.

For completeness, the behaviour of the self-energy along the \textsl{real} $\omega$ axis plays \textsl{no} vital role in the proof of the LT. For this theorem it is however vital that $\Sigma_{\upalpha\upalpha'}(\omega)$ be analytic anywhere \textsl{away from} the real axis of the complex $\omega$ plane \cite{JML61}, a fact emphasized repeatedly in Ref.~\cite{BF07a} (see \S\S2.1.2, 5, 6.2 and Appendix B herein; see in particular the discussions related to the expression in Eq.~(B.54)). The only point on the real energy axis where the behaviour of $\Sigma_{\upalpha\upalpha'}(\omega)$ is to be given attention to is $\omega = 0$ (according to the convention of Ref.~\cite{BF07a} [\S3.0.1], $\omega = \mu$, or, more precisely, $z = \mu$). Considerations in Ref.~\cite{BF07a} make explicit that insofar as the LT is concerned, from $\omega=0$ no problem arises (see \S5.3.13, in particular the discussions centred on the expression in Eq.~(5.61) of Ref.~\cite{BF07a}, as well as Appendix D herein). Interestingly, for $0 < n < N$, the singularity of the $\Sigma(\omega)$ as calculated in Ref.~\cite{DPK12} is a simple pole located at $\omega = \epsilon_0 -\mu$, where $\epsilon_0$ is defined in Eq.~(18) of Ref.~\cite{DPK12}. On identifying $\mu$ with $\mu_{\infty} = (\mu_n^-+\mu_n^+)/2$, for $U > 0$ this pole is located at $\omega=0$ only if $N$ is even \textsl{and} $n=N/2$. Remarkably, for this particular case the LT in terms of the $N_{\textsc{l}}$ presented in Eq.~(\ref{e5}) \textsl{is} satisfied.

\emph{(iii)} With reference to the data displayed in Fig.~2 of Ref.~\cite{DPK12} concerning the ``apparent doping $x_{\textsc{fs}}$ inferred from the Luttinger surface reconstruction as a function of the nominal doping $x$ in LSCO and Bi-2212'' \cite{DPK12}, we are \textsl{not} in a position to make a definitive statement about these in the absence of any information regarding the method by which $x_{\textsc{fs}}$ versus $x$ has been inferred from the experimental data of He \textsl{et al.} \cite{HZ11} and Yang \textsl{et al.} \cite{YR11}. For completeness, insofar as the LT is concerned, in these references only very brief remarks are to be found, which furthermore are suggestive of qualitative \textsl{agreement} with the LT: on p.~7 of Ref.~\cite{HZ11} one reads: ``the $k_{\textsc{f}}$ separation shows a systematic increase with doping, which is consistent with the increase of Luttinger's volume of the FS [Fermi surface] and the corresponding shift of the node position away from $k_{\textsc{af}}$'' ($k_{\textsc{af}}$ is defined in the caption of Fig.~1 of Ref.~\cite{HZ11}), and in Ref.~\cite{YR11}: ``One does not need to invoke discontinuous Fermi ``arc''s to describe the FS of underdoped Bi2212 and Luttinger's sum rule, properly understood, is seen to still approximately stand''.

In spite of the above observations, two remarks are in place. Firstly, experimental determination of the `volume' of the Fermi sea is complicated by an aspect that to our best knowledge was first spelled out by Essler and Tsvelik \cite{ET05}, and Konik, Rice and Tsvelik \cite{KRT06}. In Ref.~\cite[\S2]{BF07a} we have highlighted the problem and remarked that the Fermi sea of a metallic GS not being necessarily a \textsl{closed} set of $\bm{k}$ points, knowledge of a Fermi \textsl{surface} is in general \textsl{not} sufficient for determining the corresponding Fermi sea (see the discussions following Eq.~(2.7) in \S2.1 of Ref.~\cite{BF07a}).

Secondly, in \S6.4 of Ref.~\cite{BF07a} we have exposed the mechanism by which in the case of strongly-correlated metals (for which the single-particle momentum distribution function $\mathsf{n}_{\sigma}(\bm{k})$, pertaining to particles with spin index $\sigma$, strongly deviates from a unit-step function that is characteristic of the GSs of non-interacting fermions), some regions of the $\bm{k}$ space that are in reality external to the underlying Fermi sea, can be mistakenly identified as regions internal to this sea. As discussed in detail in Ref.~\cite[\S6.4]{BF07a}, this mechanism underlies the erroneous observation by Gr\"ober, Eder and Hanke \cite{GEH00} regarding the LT (see in particular Fig.~12 in Ref.~\cite{GEH00} and note that this figure displays the ``measured'' volume of the Fermi sea versus the \textsl{hole} concentration $1-n$, defined in relation to the half-filled, $n=1$, state of the single-band Hubbard Hamiltonian -- here, in two dimensions).

\hypertarget{item-iv}{\emph{(iv)}} With reference to an earlier work by Stanescu and Phillips \cite{SP04}, to which DPK \cite{DPK12} refer, the calculations in this work have been performed under a number of approximations (e.g. `the two-site approximation'), spelled out in \S\hspace{0.0pt}II of Ref.~\cite{SP04} (see specifically pages 5 and 6 herein). Aside from this, the computations reported in Ref.~\cite{SP04} correspond to the $\omega$-dependent functions that have been ``discretized on a grid of $N = 8192$ points from $\omega_{\textrm{min}} = -20\hspace{0.5pt}t$ to $\omega_{\textrm{max}} = 20\hspace{0.5pt}t$'', with ``all the convolutions involved in the calculation of self-energies''  having been evaluated through ``using a fast Fourier transform algorithm.'' \cite{SP04}. Here $t$ denotes the nearest-neighbour hopping integral, and $N$ should not be confused with the $N$ of the $\textrm{SU}(N)$ model. Further, ``The procedure converges for temperatures above $T = 0.02 t$ at finite doping and $T = 0.08\hspace{0.5pt}t$ at half-filling, although convergence problems occurred below $T = 0.1\hspace{0.5pt}t$.'' \cite[p.~9]{SP04}. The Fermi-surface geometries displayed in Fig.~18 of Ref.~\cite{SP04} correspond to band fillings $n=0.3$, $0.668$, $0.791$ and $0.97$, and $U = 0$, $8\hspace{0.5pt}t$ and $1000\hspace{0.5pt}t$.

In Sec.~\ref{s2b2} of the present Comment we have discussed the adverse effect of non-zero temperatures for the validity of the LT. In the following we shall therefore focus on other aspects that unquestionably render the observations in Ref.~\cite{SP04} regarding the LT unreliable.

In \S6.3 of Ref.~\cite{BF07a} we have extensively discussed the detrimental consequences of using insufficiently large cut-off energies / frequencies (denoted by $-E$ and $E$ in Ref.~\cite{BF07a}) for the self-energy and in particular the LT. In doing so, we have concentrated on some relevant works by Schmalian \emph{et al.} \cite{SLGB96a,SLGB96b}. In the calculations reported in these references, $E=30\hspace{0.5pt}t$, where $t = 0.25$\hspace{0.5pt}eV, and $N = 4096$. Clearly, the cut-off frequencies $\omega_{\textrm{min}}$ and $\omega_{\textrm{max}}$ in the calculations by Stanescu and Phillips \cite{SP04} are of the same order of magnitude as respectively $-E$ and $E$. The lowest temperature considered in Refs.~\cite{SLGB96a,SLGB96b} amounts to $T = 63$\hspace{0.5pt}K, which, making use of the relationship $1$\hspace{0.5pt}eV $\approx 1.16 \times 10^4$\hspace{0.5pt}K, amounts to approximately $0.022\hspace{0.5pt}t$. The details underlying the numerical results presented in Ref.~\cite{SP04} are not as exhaustively described as those underlying the numerical results presented in Refs.~\cite{SLGB96a,SLGB96b}. It is therefore not possible for us to be as specific with regard to the former numerical results as we have been with regard to the latter ones in Ref.~\cite[\S6.3]{BF07a}. However, from the data for the density of the single-particle states (DOS) as presented in Fig.~11 of Ref.~\cite{SP04}, one can infer that a combination of insufficiently small value of $T$ and insufficiently large value of $\vert\omega_{\textrm{min}}\vert = \omega_{\textrm{max}}$ plays a significant role in the violation of the LT as observed in Ref.~\cite{SP04}, with the latter condition being the more important of the two. See Fig.~2 of Ref.~\cite{SLGB96a}, to be compared with Fig.~4 of Ref.~\cite{BF07a} (p.~107), and consider the details centred around Eqs.~(B.42) -- (B.46) in Appendix B of Ref.~\cite{BF07a}.

We note in passing that, violation of the exact property $\im[\Sigma(\bm{k},0)] \equiv 0$, $\forall \bm{k}$, is the fundamental reason underlying the observation by Maier, Pruschke and Jarrell \cite{MPJ02} of the breakdown of the LT in the case of the single-band Hubbard Hamiltonian in two dimensions and at low concentration of holes introduced into half-filled GSs \cite[\S6.3.6]{BF07a}. For the relationship between the DOS, $\sum_{\bm{k} \in \textrm{1BZ}} A(\bm{k};\omega)$, where $A(\bm{k};\omega)$ denotes the single-particle spectral function, and $\im[\Sigma(\bm{k},\omega)]$, the reader is referred to Eq.~(B.43) in Ref.~\cite{BF07a}.

The case of $U = 1000\hspace{0.5pt}t$ (in general, large values of $U/t$) as considered by Stanescu and Phillips \cite{SP04} deserves special attention, this in the light of the considerations in Appendix D of Ref.~\cite{BSS10}. With $\omega_{\textrm{min}} = -20\hspace{0.5pt}t$ and $\omega_{\textrm{max}} = 20\hspace{0.5pt}t$, the choice of $U = 1000\hspace{0.5pt}t$ takes one \textsl{artificially} to the region of Extremely Correlated (EC) limit \cite[Eq.~(D.2)]{BSS10}, whereby breakdown of the LT becomes a certainty, on account of some non-vanishing boundary terms that appropriately have not been taken account of in the derivation of the LT (since in this derivation $E=\infty$). In the light of the explicit use in Ref.~\cite[Appendix D]{BSS10} of the single-particle Green function corresponding to the atomic limit of the Hubbard Hamiltonian \cite[\S5]{JH63}, we draw the attention of the reader to the remarks following Eq.~(\ref{e10}) in the main text of the present Comment.

To summarise, the breakdowns of the LT as observed in Ref.~\cite{SP04} are to our best judgement numerical artifacts. On account of the detailed considerations in \S6.3 of Ref.~\cite{BF07a}, we are fully confident that on redoing the calculations with sufficiently large values of $-\omega_{\textrm{min}}$ and $\omega_{\textrm{max}}$ (large in comparison with $U$), the LT will prove to be valid.

\hypertarget{item-v}{\emph{(v)}} In contrast to the statement by DPK \cite[p.~4]{DPK12}, in Ref.~\cite{KP07}, \textsl{no} ``systematic deviation'' from the requirement of the LT has been observed for the \textsl{Hubbard} model. In the concluding section of Ref.~\cite{KP07}, p.~5, one reads: ``For the Hubbard model on the $\textrm{1D}$ chain and $\textrm{2D}$ square lattice, both for $t'=0$ and $t' \not=0$, we do not find a clear-cut violation [type (iv)] of the LSR [Luttinger sum rule].''

As regards the violation of LT for the \tJ model, for which this violation has indeed been observed in Ref.~\cite{KP07}, there is no \emph{a priori} reason why the LT should be valid for this model. In fact, the theoretical considerations in Ref.~\cite{BSS10} make explicit that this violation is inherent to the \tJ model. In this connection, we point out that the criterion employed in Ref.~\cite{PLS98} (while emphasizing its limitation) for determining the Fermi surface of the metallic states of the \tJ model, namely that the Fermi surface corresponding to fermions of spin index $\sigma$ were the locus of the $\bm{k}$ points on which $\mathsf{n}_{\sigma}(\bm{k}) = \frac{1}{2}$, where $\mathsf{n}_{\sigma}(\bm{k})$ denotes the underlying GS momentum distribution function, is unfounded \cite{BF03a} (see in particular the Appendix herein). The observation in Ref.~\cite{PLS98} of the violation of the LT, based on the latter defining equation for the Fermi surface, can therefore not be viewed as reliable and thus conclusive.

For completeness, as indicated earlier by a number or authors (such as Chao, Spa{\l}ek and Ole\'{s} \cite{CSO78}, Eskes \emph{et al.} \cite{EOMS94}, Dagotto \cite{ED94}, and Eskes and Eder \cite{EE96}; see also Ref.~\cite[Ch.~5]{PF03}), the strong-coupling limit of the Hubbard Hamiltonian involves in addition to the \tJ Hamiltonian some three-site terms, proportional to $t^2/U$. In identifying the \tJ Hamiltonian as the strong-coupling limit of the Hubbard Hamiltonian, one therefore discards the latter terms, a practice that has no rigorous justification away from half-filling. In this connection, we should emphasize that even at half-filling one retains an active three-site term, arising from the anti-commutation $[i S, \mathcal{H}_{t}^0]$ (in the notation of Ref.~\cite{PF03}) associated with the canonical transformation through which the strong-coupling effective Hamiltonian $\mathcal{H}_{\textrm{eff}}$ is deduced from the Hubbard Hamiltonian $\mathcal{H}$ \cite[p.~210]{PF03}. Note that in contrast to the Hubbard model for fermions for which the Fock space of each site is spanned by four states, $\vert 0\rangle$, $\vert\hspace{-1.0pt}\uparrow\rangle$, $\vert\hspace{-1.0pt}\downarrow\rangle$ and $\vert\hspace{-1.0pt}\uparrow\downarrow\rangle$, for the \tJ model the Fock space of each site is spanned by only three states, with the doubly-occupied state $\vert\hspace{-1.0pt}\uparrow\downarrow\rangle$ excluded.

% B:
\section{Discussion of a recent work by Sakai \emph{et al.} \protect\cite{SSCMHI12}}
\label{sab}
In a recent publication regarding calculation of the self-energy $\Sigma$ of the Hubbard Hamiltonian on finite clusters in two space dimensions, with the aid of the `cellular dynamical mean-field theory', Sakai \emph{et al.} \cite{SSCMHI12} have observed an indication ``strongly'' suggestive of the failure of the LT. On the basis of the observed ``cluster-size dependence of $\Sigma$ remaining around $(\pi, 0)$ and $(\pi, \pi)$'', making it ``difficult to make definitive statements'', the authors have postponed their definite verdict on the violation or otherwise of the LT until after larger clusters will have been studied \cite[\S\hspace{0.0pt}III.C]{SSCMHI12}.

A noteworthy aspect of the computational results by Sakai \emph{et al.} \cite{SSCMHI12} is that their calculated self-energy is $\Sigma(\bm{k},i\omega_0)$, where $\omega_0 = \pi/\beta$, the Matsubara frequency $\omega_m \doteq (2 m +1)\pi/\beta$ corresponding to $m=0$; the temperature $T = 1/\beta$ in the calculations of Ref.~\cite{SSCMHI12} is chosen to be equal to $0.06\hspace{0.5pt}t$, where $t$ denotes the nearest-neighbour hopping integral (other parameters in these calculations are $t'= -0.2\hspace{0.5pt} t$ and $U = 8\hspace{0.5pt}t$, where $t'$ is the next-nearest-neighbour hopping integral, and $U$ the on-site interaction energy). Partly, however not essentially as we shall discuss below, as a result of $\omega_0$ not being infinitesimally small, the imaginary part of $\Sigma(\bm{k},i\omega_0)$, as depicted in the right-most panels of Figs.~6 and 7 in Ref.~\cite{SSCMHI12}, is non-vanishing (in fact, it is unusually large - see the following paragraph). This is \textsl{inadmissable} from the perspective of the LT, since for the exact self-energy in the zero-temperature limit one has $\im[\Sigma(\bm{k},0)] \equiv 0$, $\forall \bm{k}$ (see \S2.1.2 and Appendix B in Ref.~\cite{BF07a}; compare also with the numerical results for the real and imaginary parts of the self-energy in Fig.~1 of Ref.~\cite{ZSS95} and note that these results correspond to $t=1$ and $U = 2$). See the discussions under point \hyperlink{item-iv}{$(iv)$} in appendix \ref{saa}.

On general grounds, one expects the variation of $\im[\Sigma(\bm{k},i\omega_0)]$ as function of $\bm{k}$ to be on the scale of $\omega_0$ (cf. Eqs.~(B.55) -- (B.64) in Ref.~\cite{BF07a}), which in the case at hand is approximately equal to $0.19\hspace{0.5pt}t$, in contrast to $\re[\Sigma(\bm{k},i\omega_0)]$ whose variation is expected to be on the scale of, say, $U$. The numerical results presented in Figs.~6 and 7 of Ref.~\cite{SSCMHI12} show that for $\bm{k}$ varying over the underling $\bm{k}$-space, both $\re[\Sigma(\bm{k},i\omega_0)]$ and $\im[\Sigma(\bm{k},i\omega_0)]$ vary over intervals whose \textsl{widths} are of the order of $U$. This behaviour is inexplicable to us. Hence, it is essentially ruled out that use of $\Sigma(\bm{k},i\omega_0)$, instead of $\Sigma(\bm{k},0)$, could be the reason for the numerical results in Ref.~\cite{ZSS95} strongly suggesting violation of the LT. The employed computer codes need to be inspected for error(s).

% C:
\section{A Reply to two Replies by Kokalj and Prelov\v{s}ek \protect\cite{KP10}}
\label{sac}
In this appendix we respond to two Replies by Kokalj and Prelov\v{s}ek (hereafter KP) \cite{KP10} on the Comments in Ref.~\cite{FT09} (co-authored by A.~M. Tsvelik) and Ref.~\cite{BF09}. \emph{Below, the numbers 1), 2), \dots refer to the itemized replies in Ref.~\cite{KP10} numbered thus.}

% C.1:
\subsection{Concerning the Comment by Farid and Tsvelik \protect\cite{FT09}}
\label{sac1}
1) The remark in Ref.~\cite{FT09} ``Neither Stanescu, Phillips and Choy [11] nor KP [12] appear to have appreciated this fundamental aspect.'', to which KP object in Ref.~\cite{KP10}, has direct bearing on the following statement made in Ref.~\cite[p.~1]{KP08a}: ``On the other hand, there are several indications that LSR might be violated within the MH [Mott-Hubbard] insulators in general.$^{8-10}$''

2) --

3) As for the supposed ``claim'' made in Ref.~\cite{FT09}, in Ref.~\cite{KP08a} KP state: ``According to the LSR, for the spinless model at $n = 1/2$ one should generally have $k_{\textsc{l}} = \pi/2$ if the topology of the electronic band is not changed qualitatively (which could happen, e.g., for $t' > 0.5\hspace{0.5pt}t$). From Fig.~2 we note that $k_{\textsc{l}}$ is indeed near $\pi/2$; however, even \emph{without finite-size scaling} [our emphasis] a small deviation $k_{\textsc{l}} \not= \pi/2$ may be observed for $V > 4t$.''

In Ref.~\cite{FT09} we have made the following two main observations:

\emph{(i)} For $N=26$, where $N$ denotes the number of lattice sites, $\pi/2$ does \textsl{not} belong to the underlying $k$ space. For this very reason, for $N=26$ the $k_{\textsc{l}}$ corresponding to half-filling \textsl{cannot} have coincided with $\pi/2$. In this light, and ``without finite-size scaling'', $k_{\textsc{l}} \not= \pi/2$ cannot have signalled violation of the LT, contradicting the explicit claim to the contrary by KP \cite{KP08a,KP10} (see above).

Although indicated earlier \cite{FT09}, we emphasize that the LT (or the Luttinger sum rule, LSR) states that $N_{\textsc{l}} = N_{\textrm{e}}$ \cite[Eq.~(2)]{FT09}, where $N_{\textsc{l}}$, the LN, is defined in Eq.~(1) of Ref.~\cite{FT09}, and $N_{\textrm{e}}$ denotes the number of particles in the GS under consideration. From Fig.~2 of Ref.~\cite{KP08a} one clearly observes that for all values of $V/t$ shown (barring for the moment the value of $V/t=8$ on which KP focus in their pertinent Reply \cite{KP10}), at exactly $13$ $k$-points, out of the total of $26$ available $k$-points, one has $G(k, \mu) > 0$, in perfect agreement with the LT, given the fact that the GSs under consideration are half-filled.

We remark that from Fig.~2 of Ref.~\cite{KP08a} it appears that the solid line (drawn in red), presented as depicting the locus of $k_{\textsc{l}}$ as a function of $V/t$, is determined by taking the zero of the straight line connecting the values of $G(k,0)$, corresponding to a given value of $V/t$, at $k = 6\pi/13$ and $k= 7\pi/13$. Since the centre of the solid dot representing the value of the $G(k,0)$ corresponding to $V/t=8$ and $k = 7\pi/13$, denoted by $k_1$ in Ref.~\cite{KP10}, is visibly to the right of the middle of the red solid line depicting $k_{\textsc{l}}$, we conclude that the ``$G(k_1,\omega=0) \sim 0$'' of Ref.~\cite{KP10} in fact represents a small however \textsl{negative} value for $G(k_1,\omega=0)$, conform the requirement of the LT. Be it as it may, we feel frustrated at the fact that KP in their pertinent Reply \cite{KP10} do not disclose the actual numerical value of $G(k_1,\omega=0)$ and only suffice to present the relationship ``$G(k_1,\omega=0) \sim 0$''. It should be noted however that in the absence of any published details regarding the accuracy of the numerical results presented in Ref.~\cite{KP08a}, even a positive but sufficiently small value of $G(k_1,\omega=0)$ \textsl{cannot} be taken as signifying failure of the LT.

As for the numerical results corresponding to $N=30$, to which KP refer in Ref.~\cite{KP10}, since these are not presented in Ref.~\cite{KP08a} (nor are they in public domain), our Comment \cite{FT09} could by no means have direct bearing on them. Further, as emphasized in Ref.~\cite[\S2.5]{BF07a}, for finite systems the $k$ points at which $G(k,0) \sim 0$ have to be treated with care in taking the zero-temperature limit of the expression for the LN $N_{\textsc{l}}$.

\emph{(ii)} In the thermodynamic limit, the GS of the half-filled one-dimensional $t$-$t'$-$V$ model for small values of $t'/t$ and $V > V_{\textrm{c}}(t,t')$, where $V_{\textrm{c}}(t,t') \sim 2t$ for $t'/t \to 0$, is a charge-density-wave (CDW) state and the gap in its single-particle excitation spectrum is \textsl{not} a MH gap, but an ordinary gap arising from translational symmetry breaking \cite{FT09}. In the light of this observation, in \S\hspace{0.0pt}IV of Ref.~\cite{FT09} we have explicitly shown that in the insulating phase of the model under consideration the LT trivially holds.

4) The toy model employed in Ref.~\cite{FT09} is \textsl{not} essential for the observations in this work. It is merely intended to show that for insufficiently large values of $N$, finite-size scaling can yield very inaccurate values for $k_{\textsc{l}}$.

5) The argument presented by KP in Ref.~\cite{KP10} only further discredits their use of finite-size scaling method for arriving at a conclusion with regard to the validity or otherwise of the LT for the $t$-$t'$-$V$ model in the thermodynamic limit.

Regarding the statement in Ref.~\cite{KP10}, that ``Also it should be pointed out that the long-range order does not necessarily invalidate the LSR but merely needs a redefinition of the latter as correctly pointed out in the Comment'', in \S\hspace{0.0pt}IV of Ref.~\cite{FT09} we have explicitly dealt with this very redefinition, due to Luttinger \cite[Eq.~(95)]{JML60}.

% C.2:
\subsection{Concerning the Comment by Farid \protect\cite{BF09}}
\label{sac2}
The statement by KP in Ref.~\cite{KP10} that we merely on ``p.~3 (after Eq.~(28))" of Ref.~\cite{BF09} had `claimed' ``fallacy in the reasoning'' of Ref.~\cite{KP08b}, suggests that we must not have been sufficiently clear and exhaustive in our arguments in Ref.~\cite{BF09}. Below we attempt to compensate for these shortcomings and in doing so present some new insights that we have gained since we published Ref.~\cite{BF09} in September 2009.

1) We do \textsl{not} make any assumption of the kind suggested by KP \cite{KP10} in the analysis of Ref.~\cite{BF09}. In this reference we merely argue that $\bm{k}_{\textsc{l}}^{(1)}$ \textsl{cannot} be justifiably identified with $\bm{k}_{\textsc{l}}$, thereby to arrive at a definitive conclusion with regard to the validity or otherwise of the LT. It is relevant to recall that the vector $\bm{k}_{\textsc{l}}^{(\nu)}$, with $\nu \in \mathbb{N}$, was in explicit form first introduced in Ref.~\cite{BF09}, and \textsl{not} in Ref.~\cite{KP08b}.

2) We do \emph{not} object to the expansion in powers of $\t{\mu} \doteq \mu -U/2$, to be distinguished from the expansion in powers of $t/U$ (see later). As a matter of fact, just immediately before Eq.~(5) of Ref.~\cite{BF09} we directly refer to the expansion in powers of $\t{\mu}$ herein as ``exact''. Further, following Eq.~(6) of Ref.~\cite{BF09} we fully specify the interval of $\t{\mu}$ over which the expansion in Eq.~(5) is valid, and \textsl{nowhere} in Ref.~\cite{BF09} do we indicate or even imply that $\t{\mu}$ may be outside the latter interval; the chemical potential $\mu$ corresponding to the $N$-particle GS of the system under consideration does \textsl{not} allow for this to be the case.

3) The problem with the considerations of Ref.~\cite{KP08b} is two-fold, fundamental and technical. Before describing these, below for convenience we consider $t_{ij} = t \not= 0$ for $i$ and $j$ nearest neighbours, and $t_{ij} = 0$ otherwise. In the cases where the latter conditions are not met, the considerations are to be based on the quantity $\Delta$, instead of $t$, where $\Delta$ is introduced through the identity $t_{ij} \equiv \Delta\hspace{0.5pt} \tau_{ij}$, in which $\sum_{j} \vert \tau_{ij}\vert^2 = 1$ \cite[Eq.~(2.4)]{HL67}.

Fundamentally, insofar as the LT is concerned, $t/U = 0$ is \emph{a priori} not an appropriate limit around which to perform a \textsl{finite-order} expansion in powers of $t/U$. This is rooted in the fact that the GS of the Hubbard Hamiltonian in the atomic limit \cite[\S5]{JH63} is \textsl{not} unique (Sec.~\ref{s2a}, following Eq.~(\ref{e10})). Consequently, the LN $N_{\textsc{l}}$ as calculated in terms of the zero-temperature limit of the thermal Green function $\mathscr{G}_{\upsigma\upsigma'}(0)$, corresponding to $\mu \in (\mu_n^-,\mu_n^+)$, is in general \textsl{not} equal to $n$. From the perspective of the LT, $t/U=0$ is thus the singular point of the strong-coupling expansion of the single-particle Green function. Here $\upsigma, \upsigma' \in \{\uparrow,\downarrow\}$ denote spin indices for spin-$\frac{1}{2}$ particles.

Technically, there are two distinct aspects to be considered. Firstly, close inspection of the work by KP in Ref.~\cite{KP08b} reveals that the underlying calculations are \textsl{exactly} at the \tJ level of the strong-coupling expansion of the Hubbard Hamiltonian \cite[Ch.~5]{PF03}. It is therefore not surprising that a strict adherence to the adopted low-order formalism of Ref.~\cite{KP08b} should signal breakdown of the LT (see entry \hyperlink{item-v}{$(v)$} in appendix \ref{saa} as well as Ref.~\cite{BSS10}). For clarity, on disregarding the rest terms in the expressions for the functions $M_0^{\pm}(\bm{k})$ and $M_1^{\mp}(\bm{k})$ in Eq.~(11) of Ref.~\cite{KP08a}, one is left with functions that are specific to the \tJ model at half-filling, or the Heisenberg model (up to an unimportant additive constant due to the density-density interaction term in the \tJ Hamiltonian) \cite[\S5.1.5]{PF03}.

Secondly, assuming that the results $G^{(0,2)}(\bm{k}_{\textsc{l}}^{(1)};U/2) = O(t/U^4)$ and $G^{(1,0)}(\bm{k}_{\textsc{l}};\mu) = O(t/U^2)$, deduced by KP \cite{KP10}, are exact (see Sec.~\ref{sac2.1}), whereby
\begin{equation}
\frac{G^{(0,2)}(\bm{k}_{\textsc{l}}^{(1)};U/2)}{G^{(1,0)}(\bm{k}_{\textsc{l}};\mu)} = O\Big(\frac{1}{U^2}\Big),
\label{ec1}
\end{equation}
the asymptotic expression in Eq.~(16) of Ref.~\cite{BF09} for $k_{\textsc{l}}^{(\nu)}$ in terms of $k_{\textsc{l}}$ becomes inapplicable (for the considerations at hand, $\nu=1$). This follows from the fact that by the same reasoning that $G^{(1,0)}(\bm{k}_{\textsc{l}};\mu) = O(t/U^2)$, one has $G^{(m,0)}(\bm{k}_{\textsc{l}};\mu) = O(t/U^2)$ \cite[Eq.~(4)]{BF09} for any finite value of $m$, Sec.~\ref{sac2.1}. As a result, so long as $\vert k_{\textsc{l}}^{(\nu)}-k_{\textsc{l}}\vert$ is not sufficiently small, at the largest of the order of $(t/U)^2$ (we assume the lattice constant $a=1$), the sum with respect to $m$ in Eq.~(15) of Ref.~\cite{BF09} \textsl{cannot} be approximated by its summand corresponding to $m=1$.

With $G^{(0,2)}(\bm{k}_{\textsc{l}}^{(1)};U/2) = O(t/U^4)$ \cite{KP10}, Sec.~\ref{sac2.1}, one is thus led to the conclusion that, on account of Eq.~(12) in Ref.~\cite{BF09}, \textsl{to leading order} in $t/U$, as $t/U \to 0$, the equation to be solved for $\bm{k}_{\textsc{l}}$ is $G(\bm{k}_{\textsc{l}}^{(\nu)};\mu) = 0$, viewed as an implicit function of $\bm{k}_{\textsc{l}}$. Expressed differently, $\bm{k}_{\textsc{l}}$ is the solution of the equation $G(\bm{k};\mu) = 0$ under the subsidiary condition that to order $t/U^2$ (inclusive) the vector $\bm{k}_{\textsc{l}}^{(\nu)}$ also satisfies that latter equation. To clarify, by definition $\bm{k}_{\textsc{l}}^{(\nu)} \equiv k_{\textsc{l}}^{(\nu)} \h{\bm{n}}$ is the \textsl{exact} solution of $G_{\nu}(\bm{k};\mu) = 0$ \cite[Eq.~(13)]{BF09}. The asymptotic expression for $k_{\textsc{l}}^{(\nu)}$ in Eq.~(16) of Ref.~\cite{BF09} being deduced by equating the asymptotic expression for $G(\bm{k}_{\textsc{l}}^{(\nu)};\mu)$ on the right-hand side of Eq.~(14) of Ref.~\cite{BF09} with the aforementioned summand corresponding to $m=1$ in the expansion of $G(\bm{k}_{\textsc{l}}^{(\nu)};\mu)$ in powers of $(k_{\textsc{l}}^{(\nu)} - k_{\textsc{l}})$, it neglects an infinite number of terms, each of which through $\{G^{(m,0)}(\bm{k}_{\textsc{l}};\mu)\}$ scales to leading order like $t/U^2$, Sec.~\ref{sac2.1}, in favour of a term, i.e. $G^{(0,\nu+1)}(\bm{k}_{\textsc{l}}^{(\nu)};U/2)$, that for $t/U \to 0$ to leading order scales like $t/U^{\nu+3}$ when $\nu$ is odd and like $1/U^{\nu+2}$ when $\nu$ is even, Sec.~\ref{sac2.1}.

Following the above considerations, for a given $k_{\textsc{l}}^{(\nu)}$ \textsl{to leading order} in $t/U$ the Luttinger vector $\bm{k}_{\textsc{l}}$ is the solution of the following asymptotic equation (cf. Eq.~(15) in Ref.~\cite{BF09}):
\begin{widetext}
\begin{equation}
\sum_{m=1}^{\infty} \frac{G^{(m,0)}(\bm{k}_{\textsc{l}};\mu)}{m!}\hspace{0.6pt} (k_{\textsc{l}}^{(\nu)} - k_{\textsc{l}})^m \equiv G(\bm{k}_{\textsc{l}}^{(\nu)};\mu) - G(\bm{k}_{\textsc{l}};\mu) \sim 0.
\label{ec2}
\end{equation}
Clearly, this equation is trivially satisfied for $k_{\textsc{l}} = k_{\textsc{l}}^{(\nu)}$, the solution on which KP's \cite{KP08b,KP10} conclusion regarding the LT is based. Dividing both sides of Eq.~(\ref{ec2}) by $(k_{\textsc{l}}^{(\nu)} - k_{\textsc{l}})$, and assuming that $G^{(m,0)}(\bm{k}_{\textsc{l}};\mu) \not=0$ for $m=1,2$, one can express the resulting equation in the following appealing form:
\begin{equation}
k_{\textsc{l}} \sim k_{\textsc{l}}^{(\nu)} +2 \hspace{0.6pt} \frac{G^{(1,0)}(\bm{k}_{\textsc{l}};\mu)}{G^{(2,0)}(\bm{k}_{\textsc{l}};\mu)} \Big\{1 + \sum_{m=2}^{\infty} \frac{G^{(m+1,0)}(\bm{k}_{\textsc{l}};\mu)}{G^{(1,0)}(\bm{k}_{\textsc{l}};\mu)} \frac{(k_{\textsc{l}}^{(\nu)} - k_{\textsc{l}})^m}{(m+1)!}\Big\}.
\label{ec3}
\end{equation}
\end{widetext}
The right-hand side of this asymptotic expression depending of $\bm{k}_{\textsc{l}}$, $k_{\textsc{l}}$ is to be determined iteratively. Only for sufficiently small $\vert k_{\textsc{l}}^{(\nu)} - k_{\textsc{l}}\vert$ may one identify the $k_{\textsc{l}}$ and  $\bm{k}_{\textsc{l}}$ on the right-hand side by respectively $k_{\textsc{l}}^{(\nu)}$ and $\bm{k}_{\textsc{l}}^{(\nu)}$, and further employ the following approximate asymptotic expression:
\begin{equation}
k_{\textsc{l}} \simeq k_{\textsc{l}}^{(\nu)} + 2\hspace{0.6pt} \frac{G^{(1,0)}(\bm{k}_{\textsc{l}}^{(\nu)};\mu)}{G^{(2,0)}(\bm{k}_{\textsc{l}}^{(\nu)};\mu)}.
\label{ec4}
\end{equation}
Note that by the considerations of Sec.~\ref{sac2.1}, one has (for the lattice constant $a=1$) $G^{(1,0)}(\bm{k}_{\textsc{l}};\mu)/G^{(2,0)}(\bm{k}_{\textsc{l}};\mu) = O(1)$ and $G^{(m+1,0)}(\bm{k}_{\textsc{l}};\mu)/G^{(1,0)}(\bm{k}_{\textsc{l}};\mu) = O(1)$, $\forall m\ge 1$.

\begin{figure}[t!]
\centering
\includegraphics[angle=0, width=0.43\textwidth]{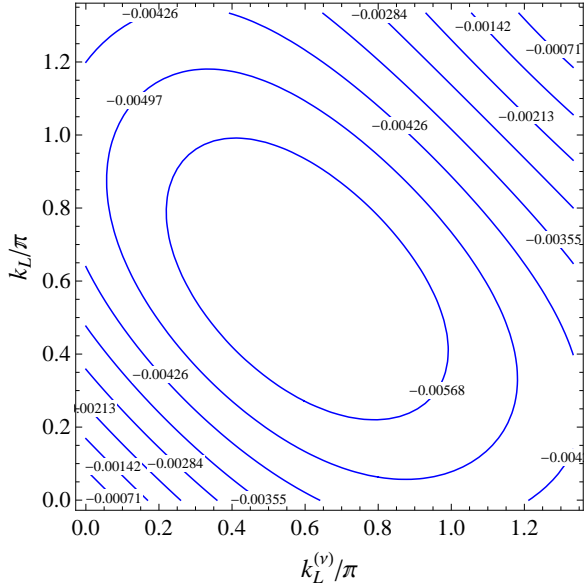}
\caption{Contour plot of the left-hand side of Eq.~(\protect\ref{ec2}) \textsl{divided by} $(k_{\textsc{l}}^{(\nu)} - k_{\textsc{l}})$ on the $k_{\textsc{l}}^{(\nu)}$-$k_{\textsc{l}}$ plane for both $\bm{k}_{\textsc{l}}^{(\nu)}$ and $\bm{k}_{\textsc{l}}$ along the positive direction of the horizontal axis of the 1BZ as specified in Eq.~(25) of Ref.~\protect\cite{BF09}, indicative of $k_{\textsc{l}} = k_{\textsc{l}}^{(\nu)}$ as being the only solution of Eq.~(\protect\ref{ec2}). As we discuss in the main text, this result was to be expected, on account of the function $G(\bm{k};\mu)$ employed here corresponding to the \tJ approximation of the strong-coupling expansion of the Hubbard Hamiltonian, Sec.~\protect\ref{sac2.1}.  The data displayed here correspond to $t=1$ and $U=40$, with the lattice constant $a$ identified with unity. }\label{fig1}
\end{figure}

In Fig.~\ref{fig1} we display a contour plot of the function on the left-hand side of Eq.~(\ref{ec2}) divided by $(k_{\textsc{l}}^{(\nu)}-k_{\textsc{l}})$. The function $G(\bm{k};\mu)$ employed in the calculations is that given in Eq.~(\ref{ec9}) for $m=0$, which is exact to order $t/U^2$ in the region $t/U\to 0$, provided that the underlying functions $M_0^{\mp}(\bm{k})$ and $M_1^{\mp}(\bm{k})$ are calculated exactly to leading order in $t/U$. The functions $M_0^{\mp}(\bm{k})$ and $M_1^{\mp}(\bm{k})$ employed in the present calculations are those presented in Eq.~(11) Ref.~\cite{KP08b}, which, as we have indicated earlier in this section, are specific to the \tJ model at half-filling, resulting in the expression for $G(\bm{k};\mu)$ as given in Eq.~(16) of Ref.~\cite{KP08b} (Eq.~(20) in Ref.~\cite{BF09}). Following this observation, it should not come as a surprise that within the present approximate framework, Eq.~(\ref{ec2}) has no other solution than $k_{\textsc{l}} = k_{\textsc{l}}^{(\nu)}$. To appreciate the significance of the shortcoming as arising from the \tJ approximation of the Hubbard Hamiltonian in the strong-coupling region, one should consider the contour lines in Fig.~\ref{fig1}, showing that a rigid upward shift of the underlying function by merely approximately $0.006$ is sufficient for producing a $k_{\textsc{l}}$ that by the required amount (from the perspective of the LT) is larger than $k_{\textsc{l}}^{(\nu)}$. See in particular the concluding paragraph of Sec.~\ref{sac2.1}. For orientation, with $\t{\mu} = t (0.1 + 6.8\hspace{0.6pt}t/U)$, for $\h{\bm{n}}$ in the positive $x$ direction of the underlying 1BZ (see Eq.~(25) in Ref.~\protect\cite{BF09}), one obtains $k_{\textsc{l}}^{(1)}/\pi \approx 0.725$, to be contrasted with $k_{\textsc{l}}/\pi \approx 0.849$, deduced by requiring satisfaction of the LT. To a good approximation, these two values clearly correspond to a point on the contour in Fig.~\ref{fig1} labeled $-0.00568$.

For completeness, and with reference to the expression in Eq.~(16) of Ref.~\cite{BF09}, we note that the order of magnitude that we have presented in Eq.~(29) of Ref.~\cite{BF09} for the coefficient of the latter equation as specialised to the case of $\nu=1$, is based on the Kramers-Kr\"onig relation and the general property of $\im[G(\bm{k};\omega \pm i 0^+)]$ for insulating GSs as $\omega$ passes through the band edges (see reference [16] in Ref.~\cite{BF09}). Further, since in Ref.~\cite{KP10} KP refer to Eq.~(29) of Ref.~\cite{BF09} (their Eq.~(2)) as ambiguous, we draw the attention of the reader to Ref.~\cite{Note4}. At any rate, we concede that the expression in Eq.~(16) of Ref.~\cite{BF09} has no root in the strong-coupling expansion of the single-particle Green function $G(\bm{k};\mu)$, in that it does not take into account the details that lead to the equalities $G^{(1,0)}(\bm{k}_{\textsc{l}};\mu) = O(t/U^2)$ and $G^{(0,2)}(\bm{k}_{\textsc{l}}^{(1)};U/2) = O(t/U^4)$. It is to be noted however that whereas the latter equality is exact, the exactness of the former one is not rigorously established. For this, see the remarks in the paragraph following that containing Eq.~(\ref{ec8}) in Sec.~\ref{sac2.1}.

\vspace{0.4cm}
% A.2.1
\subsubsection{Expressions for \texorpdfstring{$G^{(0,n)}(\bm{k};U/2)$}{} and \texorpdfstring{$G^{(m,0)}(\bm{k};\mu)$}{}}
\label{sac2.1}
In order to avoid notational confusion, unless we indicate otherwise, in this section we make use of the explicit expressions in Ref.~\cite{BF09}, where we have appropriately credited Ref.~\cite{KP08b} where credit has been due.

From the expressions in Eqs.~(8) and (11) of Ref.~\cite{BF09} one has
\begin{widetext}
\begin{equation}
G(\bm{k};\mu) = \sum_{n=0}^{\infty} (-\t{\mu})^n \Big(\frac{2}{U}\Big)^{n+1} \sum_{l=0}^{\infty} \binom{l+n}{n} \Big(\frac{2}{U}\Big)^{l}
\Big\{M_{l}^{-}(\bm{k}) - (-1)^{l+n} M_{l}^{+}(\bm{k}) \Big\},
\label{ec5}
\end{equation}
where $M_{l}^{\mp}(\bm{k})$ are defined in Eqs.~(6) and (7) of Ref.~\cite{KP08b}. Since $\t{\mu} \doteq \mu -U/2$, one has $\partial^n/\partial \mu^n \equiv \partial^n/\partial \t{\mu}^n$, $\forall n$, and further $\t{\mu} =0$ for $\mu=U/2$. Thus, from Eq.~(\ref{ec5}) one trivially obtains that (cf. Eq.~(4) in Ref.~\cite{BF09})
\begin{eqnarray}
&&\hspace{-1.0cm}G^{(0,n)}(\bm{k};U/2) = (-1)^n \Big(\frac{2}{U}\Big)^{n+1} \sum_{l=0}^{\infty} \frac{(l+n)!}{l!} \Big(\frac{2}{U}\Big)^{l}  \Big\{M_{l}^{-}(\bm{k}) - (-1)^{l+n} M_{l}^{+}(\bm{k}) \Big\} \nonumber\\
&&\hspace{-0.2cm}\sim (-1)^n n! \Big(\frac{2}{U}\Big)^{n+1} \Big\{M_{0}^{-}(\bm{k}) - (-1)^n M_{0}^{+}(\bm{k}) \Big\}
= n! \Big(\frac{2}{U}\Big)^{n+1} \times \left\{ \begin{array}{ll} 2 \b{n}_{\bm{k}s} - 1, & n = \text{even},\\ \\
1, & n= \text{odd},\end{array}\right.\;\, \text{for}\;\, \frac{t}{U} \to 0,
\label{ec6}
\end{eqnarray}
\end{widetext}
where in the last expression we have employed the first two equalities in Eq.~(8) of Ref.~\cite{KP08b}. Identifying $\b{n}_{\bm{k}s}$, the GS momentum distribution function corresponding to particles with spin index $s \in \{\uparrow,\downarrow\}$, with that corresponding to the Heisenberg model \cite[Eq.~(31)]{EOMS94} \cite[Eq.~(11)]{KP08b} \cite[Eq.~(5.45)]{PF03}, one arrives at the result $2 \b{n}_{\bm{k}s} - 1 = O(t/U)$ \cite{Note3}, specifically for $\bm{k} = \bm{k}_{\textsc{l}}^{(1)}$. Hence the equality $G^{(0,2)}(\bm{k}_{\textsc{l}}^{(1)};U/2) = O(t/U^4)$ as presented in Ref.~\cite{KP10}.

Making use of the relationship in Eq.~(3) of Ref.~\cite{BF09} and defining, in analogy with the $G^{(m,n)}(\bm{k};\mu)$ in Eq.~(4) of Ref.~\cite{BF09},
\begin{equation}
M_{l}^{\mp\hspace{0.5pt}(m)}(\bm{k}) \doteq \frac{\partial^m}{\partial k^m} M_{l}^{\mp}(\bm{k}),\; m = 0,1,2, \dots,
\label{ec7}
\end{equation}
from the expression in Eq.~(\ref{ec5}) one deduces that
\begin{widetext}
\begin{equation}
G^{(m,0)}(\bm{k};\mu) = \frac{2}{U} \sum_{n=0}^{\infty} (-\t{\mu})^n \sum_{l=n}^{\infty} \binom{l}{n} \Big(\frac{2}{U}\Big)^{l} \Big\{M_{l-n}^{-\hspace{0.5pt}(m)}(\bm{k}) - (-1)^{l} M_{l-n}^{+\hspace{0.5pt}(m)}(\bm{k}) \Big\}.
\label{ec8}
\end{equation}
\end{widetext}
Considering the expression for $\t{\mu} \doteq \mu - U/2$ as presented in Ref.~\cite{KP08b}, namely $\t{\mu} \sim -(0.19 \pm 0.1)\hspace{0.5pt}t + 6.8\hspace{0.5pt}t^2/U$ (cf. Eq.~(27) in Ref.~\cite{BF09}), one observes that $G^{(m,0)}(\bm{k}_{\textsc{l}};\mu)$ is a non-trivial function of $t/U$ and $t$.

Some technical remarks concerning the equality in Eq.~(\ref{ec8}), for $m\ge 1$, are in place. This equality has been deduced by $m$ times commuting $\partial/\partial k$ with \textsl{infinite} sums with respect to $n$ and $l$ in the expression for $G(\bm{k};\mu)$ (this expression is up to a trivial re-indexing of the summation with respect to $l$, similar to that in Eq.~(\ref{ec5})). For general $k$-dependent summands, this is admissible only if the sums with respect to $n$ and $l$ on the right-hand side of Eq.~(\ref{ec8}) are \textsl{uniformly} convergent for $k$ in a neighbourhood of the $k$ of interest, such as $k = k_{\textsc{l}}$ (assuming that these summands are \textsl{continuous} functions of $k$ throughout the last-mentioned neighbourhood and that the relevant sums corresponding to $G^{(m-1,0)}(\bm{k};\mu)$ are \textsl{convergent}). For the cases where the summands of the sums with respect to $n$ and $l$ corresponding to the function $G^{(m-1,0)}(\bm{k};\mu)$, $m\ge 1$, are \textsl{analytic} functions of $k$ in a neighbourhood of the $k$ of interest, such as $k = k_{\textsc{l}}$, the \textsl{uniform} convergence of these sums for $k$ in the latter neighbourhood suffices for commuting $\partial/\partial k$ with them, whereby $G^{(m,0)}(\bm{k};\mu)$ is expressible as in Eq.~(\ref{ec8}) \cite[\S\S4.7,5.3]{WW62}, \cite[\S46]{TJIAB65}. We note that the above-mentioned sums with respect to $n$ being power series, they are convergent in the region specified in the text following Eq.~(6) in Ref.~\cite{BF09}. In fact, in this region these sums converge \textsl{uniformly} with regard to the variable $\t{\mu}$ \cite[\S3.7]{WW62}, \cite[\S\S50-52]{TJIAB65}. As regards continuity of $G^{(m,0)}(\bm{k};\mu)$, $m\ge 0$, in its dependence on $k$, see Ref.~\cite[\S53]{TJIAB65}. We note in passing that it is in general not permissible to differentiate \textsl{asymptotic series} \cite[\S8.31]{WW62}.

From the expression in Eq.~(\ref{ec8}) one infers that the lowest-order contributions to $G^{(m,0)}(\bm{k};\mu)$ for $t/U \to 0$ arise from the terms corresponding to $(n,l)$ equal to $(0,0)$, $(0,1)$ and $(1,1)$. For $t/U \to 0$ one thus has
\begin{widetext}
\begin{equation}
G^{(m,0)}(\bm{k};\mu) \sim \frac{2}{U} \Big\{ M_0^{-\hspace{0.5pt}(m)}(\bm{k}) - M_0^{+\hspace{0.5pt}(m)}(\bm{k})\Big\} + \Big(\frac{2}{U}\Big)^2 \Big\{ \big[ M_1^{-\hspace{0.5pt}(m)}(\bm{k}) + M_1^{+\hspace{0.5pt}(m)}(\bm{k})\big] -  \big[M_0^{-\hspace{0.5pt}(m)}(\bm{k}) + M_0^{+\hspace{0.5pt}(m)}(\bm{k})\big]\hspace{0.6pt} \t{\mu} \Big\}.
\label{ec9}
\end{equation}
\end{widetext}
Making use of the expressions in Eq.~(11) of Ref.~\cite{KP08b}, one arrives at the result $G^{(1,0)}(\bm{k}_{\textsc{l}};\mu) = O(t/U^2)$, presented in Ref.~\cite{KP10}. More generally, one has $G^{(m,0)}(\bm{k}_{\textsc{l}};\mu) = O(t/U^2)$ for any finite value of $m$, including $m=0$. Clearly, the validity of this result for $m\ge 1$ is dependent on that of the expression in Eq.~(\ref{ec8}). For this, see the paragraph following the latter equation.

We point out that since $M_0^{-}(\bm{k}) + M_0^{+}(\bm{k}) \equiv 1$ for \textsl{all} $\bm{k}$, it follows that the function on the right-hand side of Eq.~(\ref{ec9}) multiplying the $\t{\mu}$ is identically vanishing for all $m \ge 1$. Consequently, $G^{(m,0)}(\bm{k};\mu)$ is to order $1/U^2$ independent of $\t{\mu}$ for all $m \ge 1$. This is a manifest shortcoming of the functions $\{G^{(m,0)}(\bm{k}_{\textsc{l}};\mu)\}$ calculated to leading order in $t/U$ for use in the calculation of $k_{\textsc{l}}$ according to the expression in Eq.~(\ref{ec2}), or that in Eq.~(\ref{ec3}). In particular with reference to the latter equation, we note that in general for $f \sim a_f \varepsilon^2 + b_f \varepsilon^3$ and $g \sim a_g \varepsilon^2 + b_g \varepsilon^3$ as $\varepsilon\to 0$, one has $f/g \sim a_f/a_g + (b_f/a_g - a_f b_g/a_g^2)\hspace{0.6pt} \varepsilon$ as $\varepsilon\to 0$. For comparison, owing to the \textsl{exact} identity $M_0^{-}(\bm{k}) + M_0^{+}(\bm{k}) \equiv 1$, $G^{(0,0)}(\bm{k};\mu) \equiv G(\bm{k};\mu)$ is to order $1/U^2$ linearly dependent on $\t{\mu}$ (cf. Eq.~(16) in Ref.~\cite{KP08b} and Eq.~(20) in Ref.~\cite{BF09}). Consequently, $k_{\textsc{l}}^{(\nu)}$, in particular for $\nu=1$, sensitively depends on the value of $\t{\mu}$.
%\hfill $\square$
\end{appendix}
%_______________________________________________________________
%

\bibliographystyle{apsrev}
%________________________
\end{document}